\ifx \undefined \onecol
\documentclass[aip,amsmath,amssymb,numerical,apl,reprint]{revtex4-1}
\else
\documentclass[aip,amsmath,amssymb,numerical,apl,preprint]{revtex4-1}
\fi

\usepackage{graphicx}% Include figure files

\newcommand{\kB}{k_B}

\newcommand{\NEPdiss}{\mathrm{NEP}^{(\mathrm{amp})}_\mathrm{diss}}

\newcommand{\Tamp}{T_\mathrm{amp}}
\newcommand{\Nqp}{N_\mathrm{qp}}
\newcommand{\nqp}{n_\mathrm{qp}}
\newcommand{\tauqp}{\tau_\mathrm{qp}}
\newcommand{\taumax}{\tau_\mathrm{max}}
\newcommand{\eopt}{\eta_\mathrm{opt}}

\newcommand{\eread}{\eta_\mathrm{read}}
\newcommand{\Pread}{P_\mathrm{read}}
\newcommand{\Pgen}{P_\mathrm{gen}}

\newcommand{\coupling}{\chi_\mathrm{c}}
\newcommand{\qpdiss}{\chi_\mathrm{qp}}

\newcommand{\Qres}{Q_r}
\newcommand{\Qi}{Q_i}
\newcommand{\Qc}{Q_c}
\newcommand{\Qiqp}{Q_{i,\mathrm{qp}}}
\newcommand{\Qimax}{Q_{i,\mathrm{max}}}

\newcommand{\nstar}{n^*_\mathrm{qp}}
\newcommand{\umm}{\mu\mathrm{m}}

\newcommand{\usm}{\mu\mathrm{s}}

\newcommand{\kifrac}{\alpha_\mathrm{sc}}
\newcommand{\Vsc}{V_\mathrm{sc}}

\newcommand{\NEP}{\mathrm{W \, Hz}^{-1/2}}
\newcommand{\res}{\mu \Omega\, \mathrm{cm}}

\begin{document}
\title{Titanium Nitride Films for Ultrasensitive Microresonator Detectors}
\begin{abstract}
\normalsize{
Titanium nitride (TiN$_x$) films are ideal for use in
superconducting microresonator detectors because:
a) the critical temperature varies with
composition ($0 < T_c < 5$~K);
b) the normal-state resistivity is large,  $\rho_n \sim 100\ \res$,
facilitating efficient photon absorption and providing a
large kinetic inductance and detector responsivity;
and c) TiN films are very hard and mechanically robust.
Resonators using reactively sputtered TiN films show remarkably
low loss ($Q_i > 10^7$) and have noise properties similar to
resonators made using other materials, while
the quasiparticle lifetimes are reasonably long, $10-200\ \mu\mathrm{s}$.
TiN microresonators should therefore
reach sensitivities well  below $10^{-19}\ \NEP$.
}
\end{abstract}
\author{Henry G. Leduc}
\affiliation{Jet Propulsion Laboratory, California Institute of
Technology, Pasadena, CA 91109}

\author{Bruce Bumble}
\affiliation{Jet Propulsion Laboratory, California Institute of
Technology, Pasadena, CA 91109}

\author{Peter K. Day}
\affiliation{Jet Propulsion Laboratory, California Institute of
Technology, Pasadena, CA 91109}

\author{Byeong Ho Eom}
\affiliation{Division of Physics, Mathematics, and Astronomy,
California Institute of Technology, Pasadena, CA 91125}

\author{Jiansong Gao}
\affiliation{National Institute of Standards and Technology, Boulder, CO, 80305}

\author{Sunil Golwala}
\affiliation{Division of Physics, Mathematics, and Astronomy,
California Institute of Technology, Pasadena, CA 91125}

\author{Benjamin A. Mazin}
\affiliation{Department of Physics, University of California, Santa Barbara CA 93106-9530}

\author{Sean McHugh}
\affiliation{Department of Physics, University of California, Santa Barbara CA 93106-9530}

\author{Andrew Merrill}
\affiliation{Department of Physics, University of California, Santa Barbara CA 93106-9530}

\author{David C. Moore}
\affiliation{Division of Physics, Mathematics, and Astronomy,
California Institute of Technology, Pasadena, CA 91125}

\author{Omid Noroozian}
\affiliation{Division of Physics, Mathematics, and Astronomy,
California Institute of Technology, Pasadena, CA 91125}

\author{Anthony D. Turner}
\affiliation{Jet Propulsion Laboratory, California Institute of
Technology, Pasadena, CA 91109}

\author{Jonas Zmuidzinas}
\affiliation{Division of Physics, Mathematics, and Astronomy,
California Institute of Technology, Pasadena, CA 91125}

\date{\today}

\maketitle

Absorption of photons with $h \nu \ge 2 \Delta \approx 3.5~k_B T_c$
in a superconductor breaks Cooper pairs into
electrons or "quasiparticles", producing a perturbation
$\delta \sigma(\omega) = \delta \sigma_1(\omega) - j \delta \sigma_2(\omega)$
of the the complex conductivity.\cite{Bluzer95,Gulian95,Sergeev96}
Such perturbations may be  readily sensed through
vector microwave measurements of lithographed microresonators,
and frequency multiplexing
enables large detector arrays.\cite{Mazin2001,Day03}
These devices are commonly known as microwave kinetic
inductance detectors, or MKIDs
because the inductive (frequency shift)
signal is considerably larger
($\beta = \delta \sigma_2 / \delta \sigma_1 \sim 3$).
However, the dissipation signal can be more sensitive,\cite{Gao07}
especially at lower modulation frequencies,
because the resonator frequency exhibits $1/f^{1/2}$
noise\cite{Day03,Mazin04,Gao07,Barends09}
caused by a surface distribution of two-level system (TLS) fluctuators.\cite{Gao07,Kumar08,Gao08b,GaoThesis,Noroozian09}
In this letter, we discuss the requirements for
ultrasensitive MKIDs and show that they
are very well fulfilled by the measured properties of
titanium nitride (TiN) films.

For dissipation readout using the standard shunt-coupled,
forward transmission ($S_{21}$) configuration\cite{Day03},
the amplifier contribution to the noise equivalent power (NEP)
is given by
\begin{equation}
\NEPdiss = 2 \frac{\Nqp \Delta}{\eopt \tauqp}
 \sqrt{\frac{\kB \Tamp}{ \coupling \qpdiss \Pread}}\ .
 \label{eqn:NEPampa}
\end{equation}
Here $\Nqp$ is the number of quasiparticles in the detector
active volume $\Vsc$;
$\eopt \approx 0.7$ is the efficiency with which photon energy
is converted to quasiparticles;\cite{Day03}
$\tauqp$ is the quasiparticle lifetime;
$\Tamp$ is the amplifier noise temperature;
$\Pread$ is the microwave readout power absorbed by the quasiparticles;
$\coupling = 4 \Qres^2/\Qc \Qi \le 1$
is optimized by matching the coupling
and internal quality factors $\Qc = \Qi$ to give
a resonator quality factor $\Qres^{-1} = \Qc^{-1} + \Qi^{-1} = 2 \Qi^{-1}$;
and $\qpdiss = \Qi/\Qiqp \le 1$ is the fraction of the resonator's
internal dissipation that is due to resistive quasiparticle
losses ($\sigma_1$).
In terms of the microwave generator power  $\Pgen$ incident on the feedline,
$\Pread = \coupling \qpdiss \Pgen/2.$
For frequency readout, the amplifier NEP is reduced by a factor of $\beta$;
however, TLS noise may then be an issue.
Note that equation~(\ref{eqn:NEPampa}) does not include
the transmission of the optical system or the absorption efficiency of the detector.

For minimizing the amplifier NEP,
$\Pread$ should be chosen so that the microwave and optical quasiparticle generation
rates are about equal, provided that the optical loading is high enough that the quasiparticle
dissipation dominates ($\qpdiss \rightarrow 1$).
In this case, amplifier noise temperatures of order
$\Tamp \sim$ 1-10~K are sufficient to achieve the photon noise
limit in the mm/submm/far-infrared bands.\cite{GaoThesis}
In this letter we are primarily concerned with the lowest NEP values  achievable,
so we examine the opposite limit of vanishingly small optical power and a correspondingly
small quasiparticle population. Other dissipation mechanisms
(radiation, TLS, grain boundaries, etc.) will then limit the resonator quality factor to some
maximum value $\Qimax$, so $\qpdiss \propto \Nqp \rightarrow 0$.
Also, the quasiparticle lifetime is observed\cite{Barends08,Kozorezov08,Barends09b}
to reach a maximum value $\taumax$
for densities $\nqp = \Nqp / \Vsc \lesssim 100\ \umm^{-3}$.
If thermally generated quasiparticles are made insignificant
by cooling and other sources of stray power are eliminated,\cite{Martinis09}
readout power dissipation remains as the only source of quasiparticles,
which are generated with some efficiency $\eread = \Nqp \Delta / \Pread \taumax \le 1$.
Eqn.~(\ref{eqn:NEPampa}) then yields
\begin{equation}
\NEPdiss \ge \frac{2}{\eopt} \sqrt{\frac{2 \eread N_0 \Delta^2 \Vsc \kB \Tamp }
{\coupling \kifrac S_1(\omega,T) \taumax \Qimax}}\ .
\end{equation}
Here $\kifrac \le 1$ is the kinetic inductance fraction,\cite{Day03}
$S_1$ is a dimensionless Mattis-Bardeen
factor of order unity,\cite{Mattis58,GaoThesis}
and $N_0$ is the single-spin density of states at the Fermi energy.
Thus,
$\mathcal{F} = \kifrac \taumax \Qimax/N_0 \Vsc$ is a useful figure of merit.
In addition, the gap parameter $\Delta$
plays a crucial role: $\NEPdiss \propto \Delta^2$
because
$\taumax^{-1} = \nstar R \propto \nstar \Delta^2$.
We will show that  $\mathcal{F}$ for TiN
is considerably better than for other materials
explored to date.

\ifx \undefined \onecol
\begin{figure}[b]
\includegraphics[width=8.5cm]{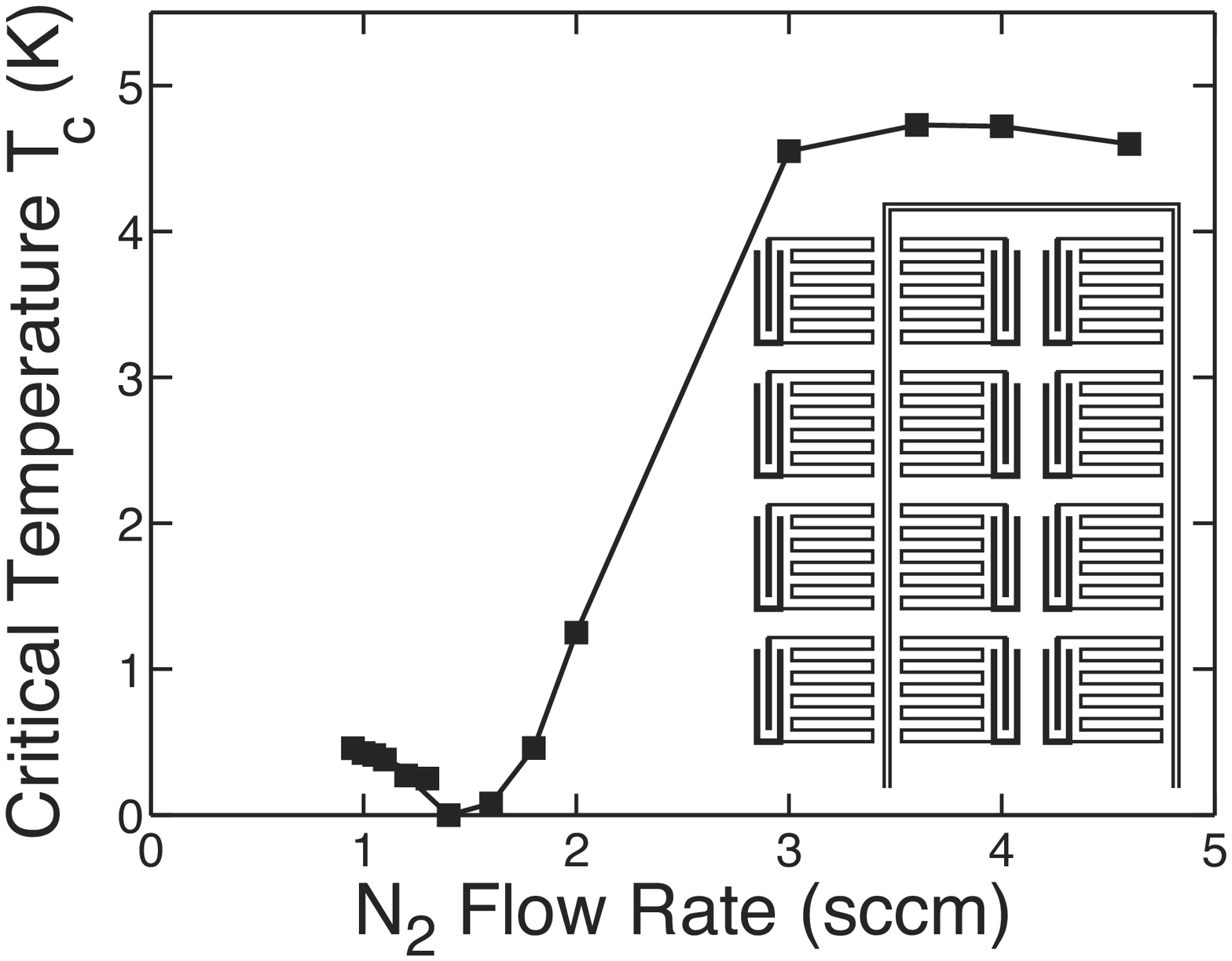}
\caption{\label{fig:fig1}
The critical temperature of reactively sputtered TiN films
as a function of the N$_2$ flow rate.
The Ar flow rate was set to 15~sccm and the total pressure to 2~mTorr.
The deposition rate was 35~nm/min using 1~kW DC power,
a 150~mm diameter Ti target, and a target to substrate distance of 15~cm.
The flow for both gases was set by thermal mass flow controllers,
while the pressure in the 17.5~L sputtering chamber was maintained
by adjusting the pump rate using a closed-loop system consisting of a
capacitance manometer,  a butterfly-type throttle valve, and a feedback
controller. The inset provides a schematic illustration of the geometry
of the $14\times16$ close-packed resonator array, with dark regions
representing TiN metallization.
}
\end{figure}
\fi

Although good resonators can be made with $T_c \approx 15$\,K
NbTiN films,\cite{Barends08b,Barends09}
lower-$T_c$ materials are needed for sensitive detectors.
We therefore studied TiN$_x$ films produced by reactive
magnetron sputtering onto ambient-temperature,
100~mm diameter, high resistivity ($> 10\, \mathrm{k}\Omega\, \mathrm{cm}$)
$\left<100\right>$ HF-cleaned silicon substrates.
The titanium sputtering target was 99.995\% pure,
and the sputtering gases (N$_2$ and Ar) were 99.9995\% pure.
As shown in Fig.~\ref{fig:fig1}, the TiN film $T_c$
is sensitive to composition.\cite{spengler78}
Microresonator structures were fabricated using deep UV projection lithography
followed by inductively coupled plasma etching using
a chlorine chemistry (BCl$_3$/Cl$_2$).
Both distributed coplanar waveguide (CPW) resonators\cite{Day03,Gao07}
as well as lumped-element resonators\cite{Doyle08}
with meandered inductors and interdigitated capacitors
(see Fig.~\ref{fig:fig2}) were produced.

For our TiN films with $0.7\ \mathrm{K} \le T_c \le 4.5\ \mathrm{K}$ and
$20\ \mathrm{nm} \le t \le 100\, \mathrm{nm}$, the normal-state resistivity
was  typically $\rho_n \approx 100\,\res$, with
$\rho_n(300~\mathrm{K}) / \rho_n(4~\mathrm{K}) \approx 1.1$.
This resistivity is similar to polycrystalline TiN films reported in
the literature but considerably higher than for single-crystal films.\cite{Johansson85}
The high resistivity (relative to Al, Ta, or Nb)
is very convenient for obtaining highly efficient far-infrared photon absorption
in lumped-element resonator structures.\cite{Doyle08}
As a consequence of
the Mattis-Bardeen relationship $L_s \approx \hbar R_s / \pi \Delta$
between the normal-state surface resistance $R_s$ and the
superconducting surface inductance $L_s$,
the large resistivity also guarantees a large kinetic inductance fraction
$\kifrac \rightarrow 1$.

\ifx \undefined \onecol
\begin{figure}[b]
\includegraphics[width=8.5cm]{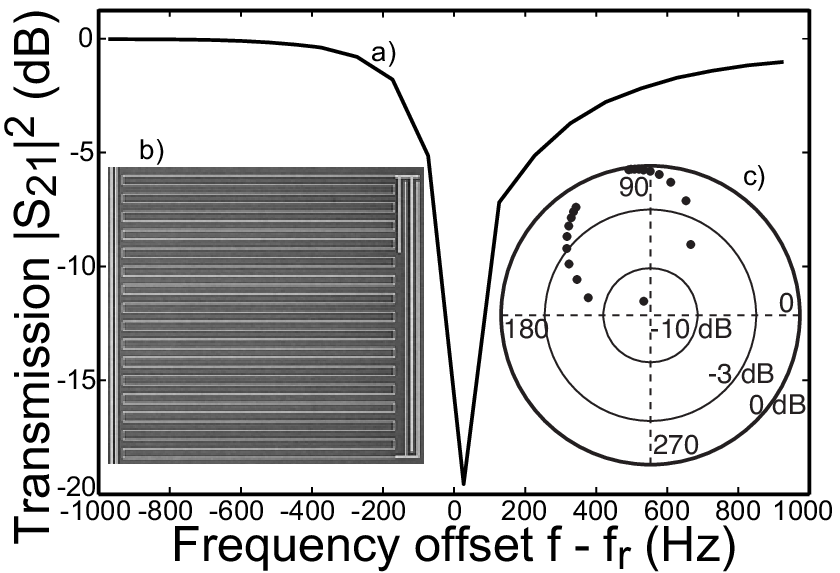}
\caption{\label{fig:fig2}
a) deep resonance measured at $T=100\,$mK and $\Pgen = -90\,$dBm
with $f_r = 1.53$~GHz,  $\Qres = 3.6 \times 10^6$, and
$\Qi = 3 \times 10^7$.
The device was a $16 \times 14$ close-packed array of lumped-element
resonators made using a $t=40$~nm TiN film with $T_c = 4.1$~K,
$R_s = 25\,\Omega$, and $L_s = 8.4$~pH.
In addition, six resonances with $\Qi > 2 \times 10^7$ were seen,
and $\ge 50$ had $\Qi > 10^7$.
The image (b) shows a single 1~mm$^2$ lumped-element resonator.
The polar $S_{21}$ plot (c) clearly shows the expected resonance loop.
}
\end{figure}
\fi

Fig.~\ref{fig:fig2}b
shows a lumped-element 1.5~GHz TiN
resonator consisting of a meandered inductor and an interdigitated capacitor (IDC),
designed to serve as a $\sim 1\ \mathrm{mm}^2$ pixel in a
$14 \times 16$ far-IR imaging array that is read out using a single coplanar
strip (CPS) feedline. The array and feedline geometry is shown schematically
in the inset to Fig.~\ref{fig:fig1}; the spacing between pixels is around 60~$\umm$.
The IDC consists of four 0.9~mm$\,\times 10\, \umm$
vertical strips with relatively large $10\, \umm$ gaps to reduce noise
and dissipation\cite{Noroozian09},
while the inductor consists of 32 1~mm$\,\times\, 5\, \umm$
horizontal strips and has $\Vsc \approx 5900\, \umm^3$ and $\kifrac \approx 0.74$.
Experiments
using a cryogenic blackbody source and a metal-mesh,
$215\,\umm$ wavelength bandpass filter
verify the basic functionality of these devices
and show that the meander is an efficient single-polarization absorber.

Although the resonators were predicted to have
$\Qc^* = 1.7 \times 10^6$,
the measured $\Qc$ values for the array show a very large
scatter $0.002 < \Qc/\Qc^* < 6$, which is
largely due to unanticipated multi-resonator modes arising
from pixel-pixel coupling.
Indeed, electromagnetic simulations show that two isolated, identically tuned, nearest-neighbor
pixels would produce symmetric and antisymmetric coupled modes with a frequency splitting
of $\sim 100$~MHz, so the interpixel coupling is much larger
than the $\sim 1$~MHz intended resonator frequency spacing.
We have since developed improved pixel designs and filled arrays with dramatically reduced
coupling; these results will be reported in a future publication.
However, the very large  accidental $\Qc$ values
have fortunately enabled a deep probe of the microwave loss of TiN.
As shown in Figs.~\ref{fig:fig2}a and \ref{fig:fig2}c, the measurements imply
$\Qimax (\mathrm{TiN}) \ge 3 \times 10^7$.
The interpretation of $\Qimax$ of the coupled modes
is secure since all resonances displayed the same
frequency vs. temperature curve and follow the Mattis-Bardeen prediction.
Furthermore, the improved uncoupled resonators also show $\Qi > 10^7$.
Regarding lower-$T_c$ material, to date our results indicate that
$\Qimax > 5 \times 10^6$ for 0.85~K TiN;
higher-$\Qc$ resonators will be needed to push this limit.
The best Al
or Nb resonators to date have $\Qimax (\mathrm{Al}) \sim 2 \times 10^6$; however,
for such high $Q$ one must generally use thick films ($t \ge 100\,$nm)
for which $\kifrac \sim 0.05$.\cite{GaoThesis} Therefore,
$\kifrac \Qimax (\mathrm{TiN}) \ge 2 \times 10^7$ whereas
$\kifrac \Qimax (\mathrm{Al, Nb}) \sim 10^5$.
We will consider the remaining factors in $\mathcal{F}$ below.

\ifx \undefined \onecol
\begin{figure}[t]
\includegraphics[width=8.5cm]{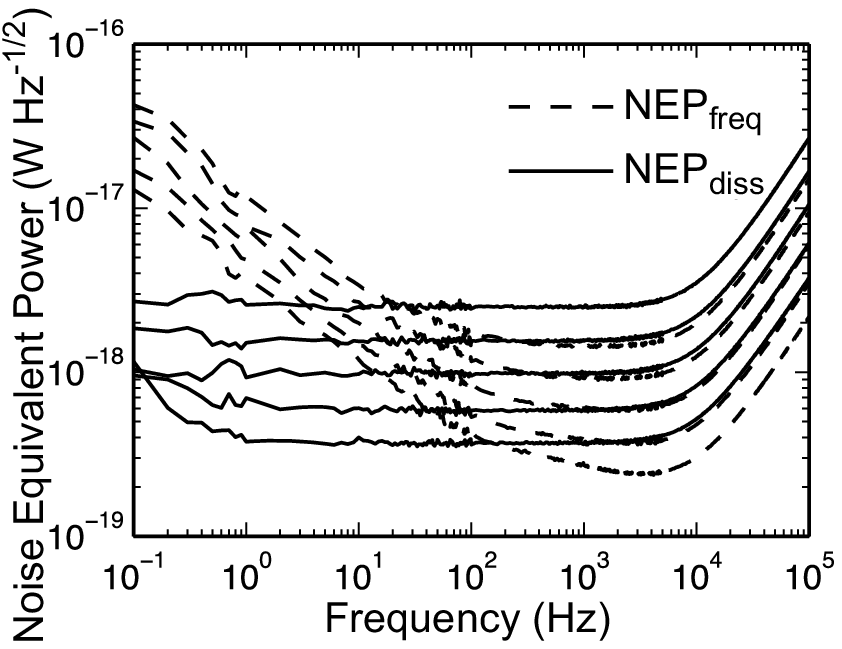}
\caption{\label{fig:fig3}
NEP for frequency readout (dashed lines) and dissipation readout (solid lines)
measured for a $t=20$~nm, $T_c = 1.1$~K TiN
CPW resonator, for readout powers
$\Pgen = -113, -109, -105, -101,$ and $ -97$~dBm (top to bottom).
The resonator center strip is $3~\mu$m wide
and 4.5~mm long, and has a $2~\mu$m gap to ground,
giving $\Vsc = 270\ \umm^3$ and $\kifrac = 0.95$.
Transmission $(S_{21})$ data measured at $T = 52$~mK and $\Pread = -93$ dBm
give $f_r = 5.380$~GHz, $\Qres = 3.2 \times 10^4$,  and $\Qi = 10^5$.
This resonance is the 3rd harmonic of the fundamental at
1.794~GHz, which was also observed but lies below the amplifier's 4-12~GHz
band.
Microwave pulse experiments and cosmic ray events
indicate $\taumax = 100\ \mu\mathrm{s}$ at $\Pgen = -109$~dBm,
consistent with $\taumax$ values seen in photon detection
experiments with other 1.1~K TiN devices.
}
\end{figure}
\fi

For ease of comparison to previous measurements, we studied the
noise of a coplanar waveguide resonator with our
standard geometry (see Fig.~\ref{fig:fig3} for details).
After correcting for the higher ($200\,\Omega$) characteristic impedance
and third-harmonic operation for the TiN CPW device\cite{GaoThesis},
the measured frequency noise
($S_f(1\,\mathrm{kHz}) \approx 3 \times 10^{-19}\, \mathrm{Hz^2 / Hz}$
at $\Pgen = -97$~dBm) may be compared to other resonators operated at
internal power $P_{\mathrm{int}} = -55$~dBm, and
is about a factor of two lower than typically seen.\cite{Gao07}
This result, along with the very similar spectral shape and power dependence,
implies that the frequency noise of the TiN device almost certainly arises from
surface TLS fluctuators.
As with other materials, no dissipation fluctuations are seen
above the cryogenic amplifier noise floor.
The corresponding electrical  NEP is
$4 \times 10^{-19}\,\NEP$ at 1~Hz
even though  $\Qc \approx 4 \times 10^4$ is quite modest.
A smaller-volume resonator with $\Qc > 10^6$ should
give an NEP in the few $10^{-20}\,\NEP$ range.

For calculating NEP, it is necessary to assume
a value for the electronic density of states $N_0$.
The results of Dridi \textit{et al.}\cite{Dridi02} are insensitive
to stochiometry over our range of interest\cite{spengler78} and
correspond to $N_0 = 8.7 \times 10^9\, \mathrm{eV}^{-1}\,\umm^{-3}$
including the electron-phonon enhancement factor
$1 + \lambda$,\cite{McMillan68,Isaev07}
or about a factor of two lower than for Al.
However, recent work\cite{Allmaier09} has indicated that
electron correlation effects in TiN may reduce $N_0$; if so this would
lower the NEP. Detection experiments with TiN resonators should help
elucidate this issue.

Another important factor is the quasiparticle lifetime.
From far-IR, UV, and X-ray photon detection experiments,
we find lifetimes of
$\taumax \approx 15\,\usm$ for $T_c = 4$~K material,
$100\,\usm$ for $T_c = 1.1$~K, and
$200\,\usm$ for $T_c = 0.8$~K,
scaling roughly as $T_c^{-2}$ as might be expected.
For $T_c = 1.1$~K, the lifetime is in the range
seen for thin Al ($t=20-40$~nm) films but
is an order of magnitude shorter
than the best thick ($t > 100$~nm) Al films.\cite{Barends09b}

Thus, the remaining factors
$\taumax/N_0 \Vsc$ contained in
$\mathcal{F}$ are about the same for
TiN, thin Al, and thick Al, to within a factor of two.
Therefore, the two orders of magnitude advantage in
$\kifrac \Qimax$ for TiN translates directly into a factor of 10
improvement in sensitivity, or for applications requiring large
sensors, an improvement of two orders of magnitude in
device area. Furthermore, the ability to reach high $\Qres$
with TiN resonators should enable very dense MKID frequency multiplexing,
and should also be of considerable interest for quantum
information and other applications.

This research was carried out in part at the Jet Propulsion Laboratory (JPL),
California Institute of Technology, under a contract with the
National Aeronautics and Space Administration.
The devices used in this work were fabricated at the JPL Microdevices
Laboratory. This work was supported in part by the NASA Science Mission
Directorate, JPL,  and the Gordon and Betty Moore Foundation.

%\bibliography{ti_nitride_c}

\begin{thebibliography}{25}%
\makeatletter
\providecommand \@ifxundefined [1]{%
 \@ifx{#1\undefined}
}%
\providecommand \@ifnum [1]{%
 \ifnum #1\expandafter \@firstoftwo
 \else \expandafter \@secondoftwo
 \fi
}%
\providecommand \@ifx [1]{%
 \ifx #1\expandafter \@firstoftwo
 \else \expandafter \@secondoftwo
 \fi
}%
\providecommand \natexlab [1]{#1}%
\providecommand \enquote  [1]{``#1''}%
\providecommand \bibnamefont  [1]{#1}%
\providecommand \bibfnamefont [1]{#1}%
\providecommand \citenamefont [1]{#1}%
\providecommand \href@noop [0]{\@secondoftwo}%
\providecommand \href [0]{\begingroup \@sanitize@url \@href}%
\providecommand \@href[1]{\@@startlink{#1}\@@href}%
\providecommand \@@href[1]{\endgroup#1\@@endlink}%
\providecommand \@sanitize@url [0]{\catcode `\\12\catcode `\$12\catcode
  `\&12\catcode `\#12\catcode `\^12\catcode `\_12\catcode `\%12\relax}%
\providecommand \@@startlink[1]{}%
\providecommand \@@endlink[0]{}%
\providecommand \url  [0]{\begingroup\@sanitize@url \@url }%
\providecommand \@url [1]{\endgroup\@href {#1}{\urlprefix }}%
\providecommand \urlprefix  [0]{URL }%
\providecommand \Eprint [0]{\href }%
\@ifxundefined \urlstyle {%
  \providecommand \doi  [0]{\begingroup \@sanitize@url \@doi}%
  \providecommand \@doi [1]{\endgroup \@@startlink {\doibase
  #1}doi:\discretionary {}{}{}#1\@@endlink }%
}{%
  \providecommand \doi  [0]{doi:\discretionary{}{}{}\begingroup
  \urlstyle{rm}\Url }%
}%
\providecommand \doibase [0]{http://dx.doi.org/}%
\providecommand \Doi [0]{\begingroup \@sanitize@url \@Doi }%
\providecommand \@Doi  [1]{\endgroup\@@startlink{\doibase#1}\@@Doi}%
\providecommand \@@Doi [1]{#1\@@endlink}%
\providecommand \selectlanguage [0]{\@gobble}%
\providecommand \bibinfo  [0]{\@secondoftwo}%
\providecommand \bibfield  [0]{\@secondoftwo}%
\providecommand \translation [1]{[#1]}%
\providecommand \BibitemOpen [0]{}%
\providecommand \bibitemStop [0]{}%
\providecommand \bibitemNoStop [0]{.\EOS\space}%
\providecommand \EOS [0]{\spacefactor3000\relax}%
\providecommand \BibitemShut  [1]{\csname bibitem#1\endcsname}%
%</preamble>
\bibitem [{\citenamefont {Bluzer}(1995)}]{Bluzer95}%
  \BibitemOpen
  \bibfield  {author} {\bibinfo {author} {\bibfnamefont {N.}~\bibnamefont
  {Bluzer}},\ }\href@noop {} {\bibfield  {journal} {\bibinfo  {journal} {J.
  Appl. Phys.},\ }\textbf {\bibinfo {volume} {{78}}},\ \bibinfo {pages} {7340}
  (\bibinfo {year} {1995})}\BibitemShut {NoStop}%
\bibitem [{\citenamefont {Gulian}\ and\ \citenamefont
  {Van~Vechten}(1995)}]{Gulian95}%
  \BibitemOpen
  \bibfield  {author} {\bibinfo {author} {\bibfnamefont {A.~M.}\ \bibnamefont
  {Gulian}}\ and\ \bibinfo {author} {\bibfnamefont {D.}~\bibnamefont
  {Van~Vechten}},\ }\href@noop {} {\bibfield  {journal} {\bibinfo  {journal}
  {Appl. Phys. Lett.},\ }\textbf {\bibinfo {volume} {{67}}},\ \bibinfo {pages}
  {2560} (\bibinfo {year} {1995})}\BibitemShut {NoStop}%
\bibitem [{\citenamefont {Sergeev}\ and\ \citenamefont
  {Reizer}(1996)}]{Sergeev96}%
  \BibitemOpen
  \bibfield  {author} {\bibinfo {author} {\bibfnamefont {A.~V.}\ \bibnamefont
  {Sergeev}}\ and\ \bibinfo {author} {\bibfnamefont {M.~Y.}\ \bibnamefont
  {Reizer}},\ }\href@noop {} {\bibfield  {journal} {\bibinfo  {journal} {Int.
  J. Mod. Phys. B},\ }\textbf {\bibinfo {volume} {{10}}},\ \bibinfo {pages}
  {635} (\bibinfo {year} {1996})}\BibitemShut {NoStop}%
\bibitem [{\citenamefont {Mazin}\ \emph {et~al.}(2002)\citenamefont {Mazin},
  \citenamefont {Day}, \citenamefont {Zmuidzinas},\ and\ \citenamefont
  {LeDuc}}]{Mazin2001}%
  \BibitemOpen
  \bibfield  {author} {\bibinfo {author} {\bibfnamefont {B.~A.}\ \bibnamefont
  {Mazin}}, \bibinfo {author} {\bibfnamefont {P.~K.}\ \bibnamefont {Day}},
  \bibinfo {author} {\bibfnamefont {J.}~\bibnamefont {Zmuidzinas}}, \ and\
  \bibinfo {author} {\bibfnamefont {H.~G.}\ \bibnamefont {LeDuc}},\ }\href@noop
  {} {\bibfield  {journal} {\bibinfo  {journal} {AIP Conf. Proc.},\ }\textbf
  {\bibinfo {volume} {605}},\ \bibinfo {pages} {309} (\bibinfo {year}
  {2002})}\BibitemShut {NoStop}%
\bibitem [{\citenamefont {Day}\ \emph {et~al.}(2003)\citenamefont {Day},
  \citenamefont {{LeDuc}}, \citenamefont {Mazin}, \citenamefont {Vayonakis},\
  and\ \citenamefont {Zmuidzinas}}]{Day03}%
  \BibitemOpen
  \bibfield  {author} {\bibinfo {author} {\bibfnamefont {P.}~\bibnamefont
  {Day}}, \bibinfo {author} {\bibfnamefont {H.~G.}\ \bibnamefont {{LeDuc}}},
  \bibinfo {author} {\bibfnamefont {B.~A.}\ \bibnamefont {Mazin}}, \bibinfo
  {author} {\bibfnamefont {A.}~\bibnamefont {Vayonakis}}, \ and\ \bibinfo
  {author} {\bibfnamefont {J.}~\bibnamefont {Zmuidzinas}},\ }\href@noop {}
  {\bibfield  {journal} {\bibinfo  {journal} {Nature},\ }\textbf {\bibinfo
  {volume} {425}},\ \bibinfo {pages} {817} (\bibinfo {year}
  {2003})}\BibitemShut {NoStop}%
\bibitem [{\citenamefont {{Gao}}\ \emph {et~al.}(2007)\citenamefont {{Gao}},
  \citenamefont {{Zmuidzinas}}, \citenamefont {{Mazin}}, \citenamefont
  {{Leduc}},\ and\ \citenamefont {{Day}}}]{Gao07}%
  \BibitemOpen
  \bibfield  {author} {\bibinfo {author} {\bibfnamefont {J.}~\bibnamefont
  {{Gao}}}, \bibinfo {author} {\bibfnamefont {J.}~\bibnamefont {{Zmuidzinas}}},
  \bibinfo {author} {\bibfnamefont {B.~A.}\ \bibnamefont {{Mazin}}}, \bibinfo
  {author} {\bibfnamefont {H.~G.}\ \bibnamefont {{Leduc}}}, \ and\ \bibinfo
  {author} {\bibfnamefont {P.~K.}\ \bibnamefont {{Day}}},\ }\href@noop {}
  {\bibfield  {journal} {\bibinfo  {journal} {Appl. Phys. Lett.},\ }\textbf
  {\bibinfo {volume} {90}},\ \bibinfo {pages} {102507} (\bibinfo {year}
  {2007})}\BibitemShut {NoStop}%
\bibitem [{\citenamefont {Mazin}(2004)}]{Mazin04}%
  \BibitemOpen
  \bibfield  {author} {\bibinfo {author} {\bibfnamefont {B.~A.}\ \bibnamefont
  {Mazin}},\ }\href@noop {} {Ph.D. thesis},\ \bibinfo  {school} {California
  Institute of Technology, Pasadena CA} (\bibinfo {year} {2004})\BibitemShut
  {NoStop}%
\bibitem [{\citenamefont {Barends}\ \emph
  {et~al.}(2009){\natexlab{a}}\citenamefont {Barends}, \citenamefont
  {Hortensius}, \citenamefont {Zijlstra}, \citenamefont {Baselmans},
  \citenamefont {Yates}, \citenamefont {Gao},\ and\ \citenamefont
  {Klapwijk}}]{Barends09}%
  \BibitemOpen
  \bibfield  {author} {\bibinfo {author} {\bibfnamefont {R.}~\bibnamefont
  {Barends}}, \bibinfo {author} {\bibfnamefont {H.~L.}\ \bibnamefont
  {Hortensius}}, \bibinfo {author} {\bibfnamefont {T.}~\bibnamefont
  {Zijlstra}}, \bibinfo {author} {\bibfnamefont {J.~J.~A.}\ \bibnamefont
  {Baselmans}}, \bibinfo {author} {\bibfnamefont {S.~J.~C.}\ \bibnamefont
  {Yates}}, \bibinfo {author} {\bibfnamefont {J.~R.}\ \bibnamefont {Gao}}, \
  and\ \bibinfo {author} {\bibfnamefont {T.~M.}\ \bibnamefont {Klapwijk}},\
  }\href@noop {} {\bibfield  {journal} {\bibinfo  {journal} {IEEE Trans. Appl.
  Supercond.},\ }\textbf {\bibinfo {volume} {19}},\ \bibinfo {pages} {936}
  (\bibinfo {year} {2009}{\natexlab{a}})}\BibitemShut {NoStop}%
\bibitem [{\citenamefont {{Kumar}}\ \emph {et~al.}(2008)\citenamefont
  {{Kumar}}, \citenamefont {{Gao}}, \citenamefont {{Zmuidzinas}}, \citenamefont
  {{Mazin}}, \citenamefont {{Leduc}},\ and\ \citenamefont {{Day}}}]{Kumar08}%
  \BibitemOpen
  \bibfield  {author} {\bibinfo {author} {\bibfnamefont {S.}~\bibnamefont
  {{Kumar}}}, \bibinfo {author} {\bibfnamefont {J.}~\bibnamefont {{Gao}}},
  \bibinfo {author} {\bibfnamefont {J.}~\bibnamefont {{Zmuidzinas}}}, \bibinfo
  {author} {\bibfnamefont {B.~A.}\ \bibnamefont {{Mazin}}}, \bibinfo {author}
  {\bibfnamefont {H.~G.}\ \bibnamefont {{Leduc}}}, \ and\ \bibinfo {author}
  {\bibfnamefont {P.~K.}\ \bibnamefont {{Day}}},\ }\href@noop {} {\bibfield
  {journal} {\bibinfo  {journal} {Appl. Phys. Lett.},\ }\textbf {\bibinfo
  {volume} {92}},\ \bibinfo {pages} {123503} (\bibinfo {year}
  {2008})}\BibitemShut {NoStop}%
\bibitem [{\citenamefont {{Gao}}\ \emph {et~al.}(2008)\citenamefont {{Gao}},
  \citenamefont {{Daal}}, \citenamefont {{Martinis}}, \citenamefont
  {{Vayonakis}}, \citenamefont {{Zmuidzinas}}, \citenamefont {{Sadoulet}},
  \citenamefont {{Mazin}}, \citenamefont {{Day}},\ and\ \citenamefont
  {{Leduc}}}]{Gao08b}%
  \BibitemOpen
  \bibfield  {author} {\bibinfo {author} {\bibfnamefont {J.}~\bibnamefont
  {{Gao}}}, \bibinfo {author} {\bibfnamefont {M.}~\bibnamefont {{Daal}}},
  \bibinfo {author} {\bibfnamefont {J.~M.}\ \bibnamefont {{Martinis}}},
  \bibinfo {author} {\bibfnamefont {A.}~\bibnamefont {{Vayonakis}}}, \bibinfo
  {author} {\bibfnamefont {J.}~\bibnamefont {{Zmuidzinas}}}, \bibinfo {author}
  {\bibfnamefont {B.}~\bibnamefont {{Sadoulet}}}, \bibinfo {author}
  {\bibfnamefont {B.~A.}\ \bibnamefont {{Mazin}}}, \bibinfo {author}
  {\bibfnamefont {P.~K.}\ \bibnamefont {{Day}}}, \ and\ \bibinfo {author}
  {\bibfnamefont {H.~G.}\ \bibnamefont {{Leduc}}},\ }\href@noop {} {\bibfield
  {journal} {\bibinfo  {journal} {Appl. Phys. Lett.},\ }\textbf {\bibinfo
  {volume} {92}},\ \bibinfo {pages} {212504} (\bibinfo {year}
  {2008})}\BibitemShut {NoStop}%
\bibitem [{\citenamefont {Gao}(2008)}]{GaoThesis}%
  \BibitemOpen
  \bibfield  {author} {\bibinfo {author} {\bibfnamefont {J.}~\bibnamefont
  {Gao}},\ }\href@noop {} {Ph.D. thesis},\ \bibinfo  {school} {California
  Institute of Technology, Pasadena CA} (\bibinfo {year} {2008})\BibitemShut
  {NoStop}%
\bibitem [{\citenamefont {{Noroozian}}\ \emph {et~al.}(2009)\citenamefont
  {{Noroozian}}, \citenamefont {{Gao}}, \citenamefont {{Zmuidzinas}},
  \citenamefont {{Leduc}},\ and\ \citenamefont {{Mazin}}}]{Noroozian09}%
  \BibitemOpen
  \bibfield  {author} {\bibinfo {author} {\bibfnamefont {O.}~\bibnamefont
  {{Noroozian}}}, \bibinfo {author} {\bibfnamefont {J.}~\bibnamefont {{Gao}}},
  \bibinfo {author} {\bibfnamefont {J.}~\bibnamefont {{Zmuidzinas}}}, \bibinfo
  {author} {\bibfnamefont {H.~G.}\ \bibnamefont {{Leduc}}}, \ and\ \bibinfo
  {author} {\bibfnamefont {B.~A.}\ \bibnamefont {{Mazin}}},\ }\href@noop {}
  {\bibfield  {journal} {\bibinfo  {journal} {AIP Conf. Proc.},\ }\textbf
  {\bibinfo {volume} {1185}},\ \bibinfo {pages} {148} (\bibinfo {year}
  {2009})}\BibitemShut {NoStop}%
\bibitem [{\citenamefont {Barends}\ \emph
  {et~al.}(2008){\natexlab{a}}\citenamefont {Barends}, \citenamefont
  {Baselmans}, \citenamefont {Yates}, \citenamefont {Gao}, \citenamefont
  {Hovenier},\ and\ \citenamefont {Klapwijk}}]{Barends08}%
  \BibitemOpen
  \bibfield  {author} {\bibinfo {author} {\bibfnamefont {R.}~\bibnamefont
  {Barends}}, \bibinfo {author} {\bibfnamefont {J.~J.~A.}\ \bibnamefont
  {Baselmans}}, \bibinfo {author} {\bibfnamefont {S.~J.~C.}\ \bibnamefont
  {Yates}}, \bibinfo {author} {\bibfnamefont {J.~R.}\ \bibnamefont {Gao}},
  \bibinfo {author} {\bibfnamefont {J.~N.}\ \bibnamefont {Hovenier}}, \ and\
  \bibinfo {author} {\bibfnamefont {T.~M.}\ \bibnamefont {Klapwijk}},\
  }\href@noop {} {\bibfield  {journal} {\bibinfo  {journal} {Phys. Rev.
  Lett.},\ }\textbf {\bibinfo {volume} {{100}}},\ \bibinfo {pages} {257002}
  (\bibinfo {year} {2008}{\natexlab{a}})}\BibitemShut {NoStop}%
\bibitem [{\citenamefont {Kozorezov}\ \emph {et~al.}(2008)\citenamefont
  {Kozorezov}, \citenamefont {Golubov}, \citenamefont {Wigmore}, \citenamefont
  {Martin}, \citenamefont {Verhoeve}, \citenamefont {Hijmering},\ and\
  \citenamefont {Jerjen}}]{Kozorezov08}%
  \BibitemOpen
  \bibfield  {author} {\bibinfo {author} {\bibfnamefont {A.~G.}\ \bibnamefont
  {Kozorezov}}, \bibinfo {author} {\bibfnamefont {A.~A.}\ \bibnamefont
  {Golubov}}, \bibinfo {author} {\bibfnamefont {J.~K.}\ \bibnamefont
  {Wigmore}}, \bibinfo {author} {\bibfnamefont {D.}~\bibnamefont {Martin}},
  \bibinfo {author} {\bibfnamefont {P.}~\bibnamefont {Verhoeve}}, \bibinfo
  {author} {\bibfnamefont {R.~A.}\ \bibnamefont {Hijmering}}, \ and\ \bibinfo
  {author} {\bibfnamefont {I.}~\bibnamefont {Jerjen}},\ }\href@noop {}
  {\bibfield  {journal} {\bibinfo  {journal} {Phys. Rev. B},\ }\textbf
  {\bibinfo {volume} {{78}}},\ \bibinfo {pages} {{174501}} (\bibinfo {year}
  {2008})}\BibitemShut {NoStop}%
\bibitem [{\citenamefont {Barends}\ \emph
  {et~al.}(2009){\natexlab{b}}\citenamefont {Barends}, \citenamefont {van
  Vliet}, \citenamefont {Baselmans}, \citenamefont {Yates}, \citenamefont
  {Gao},\ and\ \citenamefont {Klapwijk}}]{Barends09b}%
  \BibitemOpen
  \bibfield  {author} {\bibinfo {author} {\bibfnamefont {R.}~\bibnamefont
  {Barends}}, \bibinfo {author} {\bibfnamefont {S.}~\bibnamefont {van Vliet}},
  \bibinfo {author} {\bibfnamefont {J.~J.~A.}\ \bibnamefont {Baselmans}},
  \bibinfo {author} {\bibfnamefont {S.~J.~C.}\ \bibnamefont {Yates}}, \bibinfo
  {author} {\bibfnamefont {J.~R.}\ \bibnamefont {Gao}}, \ and\ \bibinfo
  {author} {\bibfnamefont {T.~M.}\ \bibnamefont {Klapwijk}},\ }\href@noop {}
  {\bibfield  {journal} {\bibinfo  {journal} {Phys. Rev. B},\ }\textbf
  {\bibinfo {volume} {79}},\ \bibinfo {pages} {020509} (\bibinfo {year}
  {2009}{\natexlab{b}})}\BibitemShut {NoStop}%
\bibitem [{\citenamefont {Martinis}\ \emph {et~al.}(2009)\citenamefont
  {Martinis}, \citenamefont {Ansmann},\ and\ \citenamefont
  {Aumentado}}]{Martinis09}%
  \BibitemOpen
  \bibfield  {author} {\bibinfo {author} {\bibfnamefont {J.~M.}\ \bibnamefont
  {Martinis}}, \bibinfo {author} {\bibfnamefont {M.}~\bibnamefont {Ansmann}}, \
  and\ \bibinfo {author} {\bibfnamefont {J.}~\bibnamefont {Aumentado}},\
  }\href@noop {} {\bibfield  {journal} {\bibinfo  {journal} {Phys. Rev.
  Lett.},\ }\textbf {\bibinfo {volume} {103}},\ \bibinfo {pages} {{097002}}
  (\bibinfo {year} {2009})}\BibitemShut {NoStop}%
\bibitem [{\citenamefont {Mattis}\ and\ \citenamefont
  {Bardeen}(1958)}]{Mattis58}%
  \BibitemOpen
  \bibfield  {author} {\bibinfo {author} {\bibfnamefont {D.~C.}\ \bibnamefont
  {Mattis}}\ and\ \bibinfo {author} {\bibfnamefont {J.}~\bibnamefont
  {Bardeen}},\ }\href@noop {} {\bibfield  {journal} {\bibinfo  {journal} {Phys.
  Rev.},\ }\textbf {\bibinfo {volume} {111}},\ \bibinfo {pages} {412} (\bibinfo
  {year} {1958})}\BibitemShut {NoStop}%
\bibitem [{\citenamefont {Barends}\ \emph
  {et~al.}(2008){\natexlab{b}}\citenamefont {Barends}, \citenamefont
  {Hortensius}, \citenamefont {Zijlstra}, \citenamefont {Baselmans},
  \citenamefont {Yates}, \citenamefont {Gao},\ and\ \citenamefont
  {Klapwijk}}]{Barends08b}%
  \BibitemOpen
  \bibfield  {author} {\bibinfo {author} {\bibfnamefont {R.}~\bibnamefont
  {Barends}}, \bibinfo {author} {\bibfnamefont {H.~L.}\ \bibnamefont
  {Hortensius}}, \bibinfo {author} {\bibfnamefont {T.}~\bibnamefont
  {Zijlstra}}, \bibinfo {author} {\bibfnamefont {J.~J.~A.}\ \bibnamefont
  {Baselmans}}, \bibinfo {author} {\bibfnamefont {S.~J.~C.}\ \bibnamefont
  {Yates}}, \bibinfo {author} {\bibfnamefont {J.~R.}\ \bibnamefont {Gao}}, \
  and\ \bibinfo {author} {\bibfnamefont {T.~M.}\ \bibnamefont {Klapwijk}},\
  }\href@noop {} {\bibfield  {journal} {\bibinfo  {journal} {Appl. Phys.
  Lett.},\ }\textbf {\bibinfo {volume} {92}},\ \bibinfo {pages} {223502}
  (\bibinfo {year} {2008}{\natexlab{b}})}\BibitemShut {NoStop}%
\bibitem [{\citenamefont {Spengler}\ \emph {et~al.}(1978)\citenamefont
  {Spengler}, \citenamefont {Kaiser}, \citenamefont {Christensen},\ and\
  \citenamefont {M\"uller-Vogt}}]{spengler78}%
  \BibitemOpen
  \bibfield  {author} {\bibinfo {author} {\bibfnamefont {W.}~\bibnamefont
  {Spengler}}, \bibinfo {author} {\bibfnamefont {R.}~\bibnamefont {Kaiser}},
  \bibinfo {author} {\bibfnamefont {A.~N.}\ \bibnamefont {Christensen}}, \ and\
  \bibinfo {author} {\bibfnamefont {G.}~\bibnamefont {M\"uller-Vogt}},\
  }\href@noop {} {\bibfield  {journal} {\bibinfo  {journal} {Phys. Rev. B},\
  }\textbf {\bibinfo {volume} {17}},\ \bibinfo {pages} {1095} (\bibinfo {year}
  {1978})}\BibitemShut {NoStop}%
\bibitem [{\citenamefont {Doyle}\ \emph {et~al.}(2008)\citenamefont {Doyle},
  \citenamefont {Mauskopf}, \citenamefont {Naylon}, \citenamefont {Porch},\
  and\ \citenamefont {Duncombe}}]{Doyle08}%
  \BibitemOpen
  \bibfield  {author} {\bibinfo {author} {\bibfnamefont {S.}~\bibnamefont
  {Doyle}}, \bibinfo {author} {\bibfnamefont {P.}~\bibnamefont {Mauskopf}},
  \bibinfo {author} {\bibfnamefont {J.}~\bibnamefont {Naylon}}, \bibinfo
  {author} {\bibfnamefont {A.}~\bibnamefont {Porch}}, \ and\ \bibinfo {author}
  {\bibfnamefont {C.}~\bibnamefont {Duncombe}},\ }\href@noop {} {\bibfield
  {journal} {\bibinfo  {journal} {J. Low Temp. Phys.},\ }\textbf {\bibinfo
  {volume} {{151}}},\ \bibinfo {pages} {530} (\bibinfo {year}
  {2008})}\BibitemShut {NoStop}%
\bibitem [{\citenamefont {Johansson}\ \emph {et~al.}(1985)\citenamefont
  {Johansson}, \citenamefont {Sundgren}, \citenamefont {Greene}, \citenamefont
  {Rockett},\ and\ \citenamefont {Barnett}}]{Johansson85}%
  \BibitemOpen
  \bibfield  {author} {\bibinfo {author} {\bibfnamefont {B.~O.}\ \bibnamefont
  {Johansson}}, \bibinfo {author} {\bibfnamefont {J.-E.}\ \bibnamefont
  {Sundgren}}, \bibinfo {author} {\bibfnamefont {J.~E.}\ \bibnamefont
  {Greene}}, \bibinfo {author} {\bibfnamefont {A.}~\bibnamefont {Rockett}}, \
  and\ \bibinfo {author} {\bibfnamefont {S.~A.}\ \bibnamefont {Barnett}},\
  }\href@noop {} {\bibfield  {journal} {\bibinfo  {journal} {J. Vac. Sci.
  Technol. A},\ }\textbf {\bibinfo {volume} {3}},\ \bibinfo {pages} {303}
  (\bibinfo {year} {1985})}\BibitemShut {NoStop}%
\bibitem [{\citenamefont {{Dridi}}\ \emph {et~al.}(2002)\citenamefont
  {{Dridi}}, \citenamefont {{Bouhafs}}, \citenamefont {{Ruterana}},\ and\
  \citenamefont {{Aourag}}}]{Dridi02}%
  \BibitemOpen
  \bibfield  {author} {\bibinfo {author} {\bibfnamefont {Z.}~\bibnamefont
  {{Dridi}}}, \bibinfo {author} {\bibfnamefont {B.}~\bibnamefont {{Bouhafs}}},
  \bibinfo {author} {\bibfnamefont {P.}~\bibnamefont {{Ruterana}}}, \ and\
  \bibinfo {author} {\bibfnamefont {H.}~\bibnamefont {{Aourag}}},\ }\href@noop
  {} {\bibfield  {journal} {\bibinfo  {journal} {J. Phys. Cond. Matt.},\
  }\textbf {\bibinfo {volume} {14}},\ \bibinfo {pages} {10237} (\bibinfo {year}
  {2002})}\BibitemShut {NoStop}%
\bibitem [{\citenamefont {McMillan}(1968)}]{McMillan68}%
  \BibitemOpen
  \bibfield  {author} {\bibinfo {author} {\bibfnamefont {W.}~\bibnamefont
  {McMillan}},\ }\href@noop {} {\bibfield  {journal} {\bibinfo  {journal}
  {Phys. Rev.},\ }\textbf {\bibinfo {volume} {167}},\ \bibinfo {pages} {331}
  (\bibinfo {year} {1968})}\BibitemShut {NoStop}%
\bibitem [{\citenamefont {{Isaev}}\ \emph {et~al.}(2007)\citenamefont
  {{Isaev}}, \citenamefont {{Simak}}, \citenamefont {{Abrikosov}},
  \citenamefont {{Ahuja}}, \citenamefont {{Vekilov}}, \citenamefont
  {{Katsnelson}}, \citenamefont {{Lichtenstein}},\ and\ \citenamefont
  {{Johansson}}}]{Isaev07}%
  \BibitemOpen
  \bibfield  {author} {\bibinfo {author} {\bibfnamefont {E.~I.}\ \bibnamefont
  {{Isaev}}}, \bibinfo {author} {\bibfnamefont {S.~I.}\ \bibnamefont
  {{Simak}}}, \bibinfo {author} {\bibfnamefont {I.~A.}\ \bibnamefont
  {{Abrikosov}}}, \bibinfo {author} {\bibfnamefont {R.}~\bibnamefont
  {{Ahuja}}}, \bibinfo {author} {\bibfnamefont {Y.~K.}\ \bibnamefont
  {{Vekilov}}}, \bibinfo {author} {\bibfnamefont {M.~I.}\ \bibnamefont
  {{Katsnelson}}}, \bibinfo {author} {\bibfnamefont {A.~I.}\ \bibnamefont
  {{Lichtenstein}}}, \ and\ \bibinfo {author} {\bibfnamefont {B.}~\bibnamefont
  {{Johansson}}},\ }\href@noop {} {\bibfield  {journal} {\bibinfo  {journal}
  {J. Appl. Phys.},\ }\textbf {\bibinfo {volume} {101}},\ \bibinfo {pages}
  {123519} (\bibinfo {year} {2007})}\BibitemShut {NoStop}%
\bibitem [{\citenamefont {Allmaier}\ \emph {et~al.}(2009)\citenamefont
  {Allmaier}, \citenamefont {Chioncel},\ and\ \citenamefont
  {Arrigoni}}]{Allmaier09}%
  \BibitemOpen
  \bibfield  {author} {\bibinfo {author} {\bibfnamefont {H.}~\bibnamefont
  {Allmaier}}, \bibinfo {author} {\bibfnamefont {L.}~\bibnamefont {Chioncel}},
  \ and\ \bibinfo {author} {\bibfnamefont {E.}~\bibnamefont {Arrigoni}},\
  }\href@noop {} {\bibfield  {journal} {\bibinfo  {journal} {Phys. Rev. B},\
  }\textbf {\bibinfo {volume} {79}},\ \bibinfo {pages} {235126} (\bibinfo
  {year} {2009})}\BibitemShut {NoStop}%
\end{thebibliography}

%merlin.mbs 2010-03-15 4.21a (PWD, AO, DPC)

\ifx \undefined \onecol
\else

\pagebreak
\begin{figure}[h]
\includegraphics[width=8.5cm]{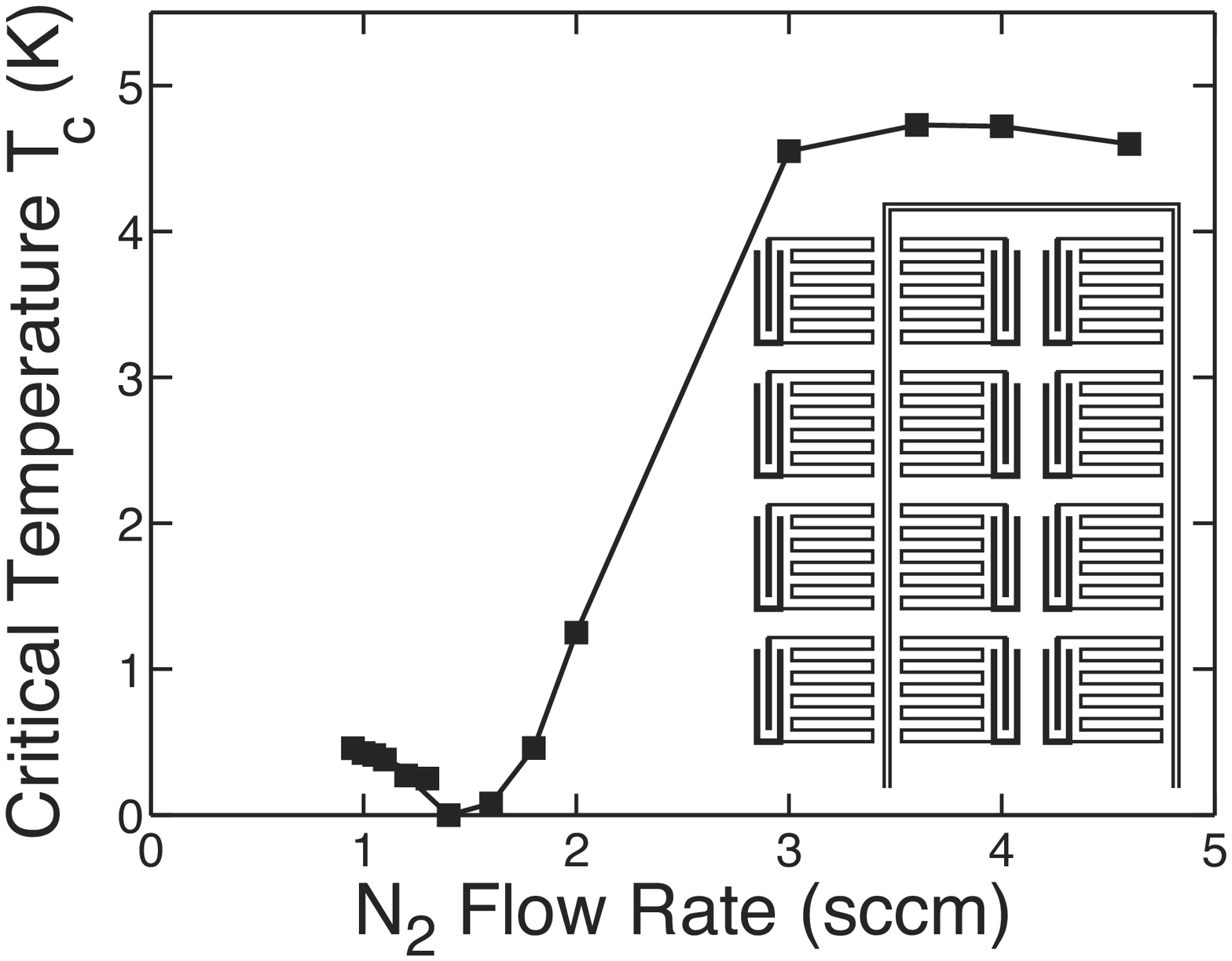}
\caption{\label{fig:fig1}
The critical temperature of reactively sputtered TiN films
as a function of the N$_2$ flow rate.
The Ar flow rate was set to 15~sccm and the total pressure to 2~mTorr.
The deposition rate was 35~nm/min using 1~kW DC power,
a 150~mm diameter Ti target, and a target to substrate distance of 15~cm.
The flow for both gases was set by thermal mass flow controllers,
while the pressure in the 17.5~L sputtering chamber was maintained
by adjusting the pump rate using a closed-loop system consisting of a
capacitance manometer,  a butterfly-type throttle valve, and a feedback
controller. The inset provides a schematic illustration of the geometry
of the $14\times16$ close-packed resonator array, with dark regions
representing TiN metallization.
}
\end{figure}

\pagebreak
\begin{figure}[h]
\includegraphics[width=8.5cm]{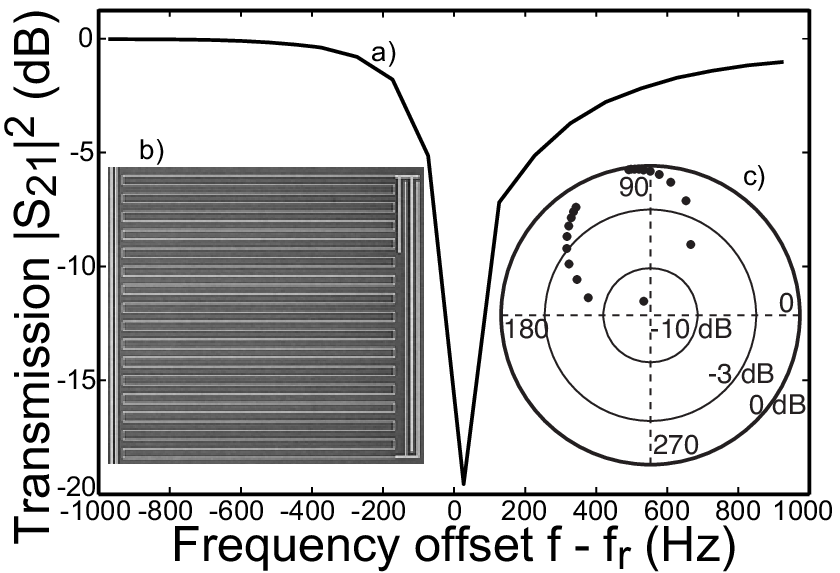}
\caption{\label{fig:fig2}
a) deep resonance measured at $T=100\,$mK and $\Pgen = -90\,$dBm
with $f_r = 1.53$~GHz,  $\Qres = 3.6 \times 10^6$, and
$\Qi = 3 \times 10^7$.
The device was a $16 \times 14$ close-packed array of lumped-element
resonators made using a $t=40$~nm TiN film with $T_c = 4.1$~K,
$R_s = 25\,\Omega$, and $L_s = 8.4$~pH.
In addition, six resonances with $\Qi > 2 \times 10^7$ were seen,
and $\ge 50$ had $\Qi > 10^7$.
The image (b) shows a single 1~mm$^2$ lumped-element resonator.
The polar $S_{21}$ plot (c) clearly shows the expected resonance loop.
}
\end{figure}

\pagebreak
\begin{figure}[h]
\includegraphics[width=8.5cm]{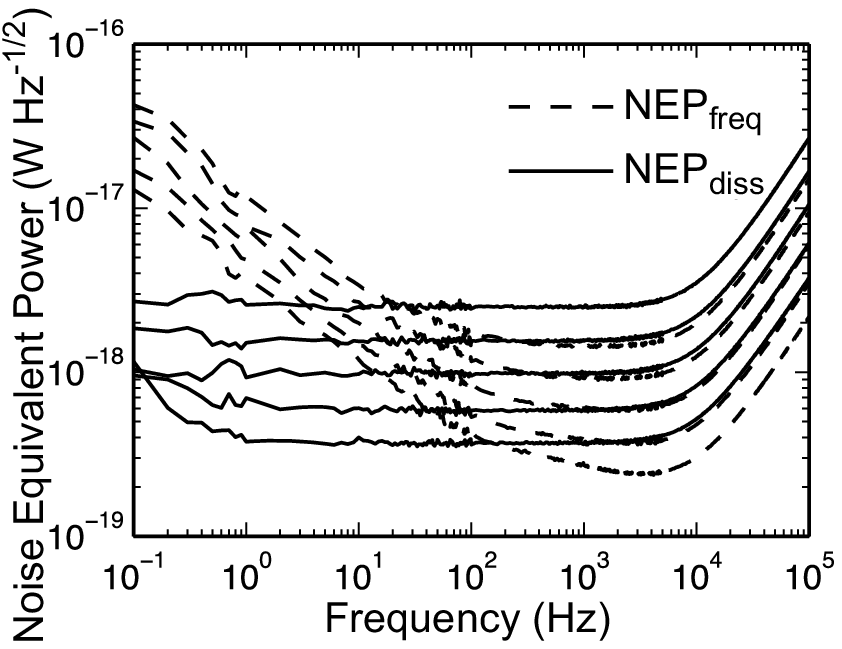}
\caption{\label{fig:fig3}
NEP for frequency readout (dashed lines) and dissipation readout (solid lines)
measured for a $t=20$~nm, $T_c = 1.1$~K TiN
CPW resonator, for readout powers
$\Pgen = -113, -109, -105, -101,$ and $ -97$~dBm (top to bottom).
The resonator center strip is $3~\mu$m wide
and 4.5~mm long, and has a $2~\mu$m gap to ground,
giving $\Vsc = 270\ \umm^3$ and $\kifrac = 0.95$.
Transmission $(S_{21})$ data measured at $T = 52$~mK and $\Pread = -93$ dBm
give $f_r = 5.380$~GHz, $\Qres = 3.2 \times 10^4$,  and $\Qi = 10^5$.
This resonance is the 3rd harmonic of the fundamental at
1.794~GHz, which was also observed but lies below the amplifier's 4-12~GHz
band.
Microwave pulse experiments and cosmic ray events
indicate $\taumax = 100\ \mu\mathrm{s}$ at $\Pgen = -109$~dBm,
consistent with $\taumax$ values seen in photon detection
experiments with other 1.1~K TiN devices.
}
\end{figure}
\fi

\end{document}